# Occupational Health Problem Network : the Exposome

Laurie Faisandier[1], Régis De Gaudemaris[1,2], Dominique J Bicout[3]

[1] Laboratoire Environnement et Prédiction de la Santé des Populations-TIMC, Université Joseph Fourier, Grenoble, Domaine de la Merci, 38706 La Tronche, France
[2] Service de Médecine et Santé au Travail, CHU Grenoble, BP 217, 38043 Grenoble cedex 09 France
[3] Biomathématiques et Epidémiologie, Laboratoire Environnement et Prédiction de la Santé des Populations-TIMC, UMR CNRS 5525 Université Joseph Fourier, Ecole Nationale Vétérinaire de Lyon, 69280 Marcy l'Etoile, France

Corresponding authors: LFaisandier@chu-grenoble.fr, d.bicout@vet-lyon.fr

In the past two decades, industrialisation, urbanisation, the booming development of the transport industry and intensive farming have all resulted in considerable changes at both an individual and societal level. The professional environment has in particular undergone technological advances with the discovery of new substances and the modernisation of manufacturing processes. With an increase in the incidence of pathologies whose root causes are poorly understood and sometimes not even known, and a growing demand by the public for the improvement of working conditions, these developments and changes raise general questions about the relationship between the environment (professional) and health.

Initiated in 2001, the Réseau National de Vigilance et de prevention des Pathologies Professionelles (RNV3P) is a computer-based network which collates information on health problems in the work place, issued by the centres of occupational diseases throughout France. The observations are reported by doctors – expert database findings grow each year by, on average, 5,000 occupational health problems. The objectives of the RNV3P are based on two main principals: the identification of professional situations that expose people to risk and the aetiology of emerging medical events associated with problems of health at work. With a substantial number of observations reported, the RNV3P constitutes an essential resource in the prospective epidemiological surveillance of occupational diseases.

An occupational-health problem (OHP) is seen in the situation where a patient presents with a principal pathology associated with a cascade of up to 5 noxious agents that are related to the place of work, the profession (or job) and the patient's particular activity. A degree of responsibility is attributed to each noxious substance and reflects the intensity of the causal-effect relationship of the substance on the presentation of disease. Thus two OHPs may differ either by principle pathology and/or an occupational exposure defined by noxious agent, profession or sector of activity. Likewise, several patients can present with exactly the same OHP.

In the face of so many individual factors and complex professional situations where exposures of diverse origins and variable intensity combine over time to have an effect on health, it may be helpful to investigate characteristics or traits which gather or separate OHPs. In effect, many diseases are of multifactorial origin due to a prolonged and regular exposure to a range of noxious products. Because of this, several distinct OHPs are likely to have in common similar professional situations. Consequently, it would be convenient to develop approaches capable of analysing all OHPs in a global manner for the study of pathology-exposure relationships. For this reason, we have developed the concept of "exposome", defined as a relational network of OHPs having in common an element at least of an occupational exposure: the noxious substances, job or sector of activity [1]. Thus, the RNV3P database is structured by groups of OHPs sharing characteristics in common. In this way, exposomes represent a tripartite occupational exposure – pathology network, founded upon presentation of disease.

Figure 1 illustrates the structure encased in the network with the exposomes at different levels of organisation, indicating the relational complexity between OHPs. Each link in the exposome represents an OHP and the size of the link is proportional to the number of identical OHPs. Two links are connected when they have at least one element of occupational exposure in common. The number of connections between links indicates both the number of different exposures (multi-exposures) and the diversity of connections from the number of different neighbouring OHPS. In this way, the organisation of OHPs in the exposome allows identification of groups of OHPs that share similar occupational exposures and inter-connected or linked OHPs belonging to several groups. This exposome network is then used to conduct an epidemiological survey by the programmed prospective surveillance of OHP groups (or associations pathology x occupational exposures) identified and of a prospective surveillance by the detection of emerging events as new links and/or new connections appear [2]. By definition, these exposomes should also enable study of synergistic relationships of occupational exposures on the OHPs and to construct hypotheses on the aetiology of pathologies.

The term exposome has already been used in the literature. In effect, just as for health risk analysis, where the toxicological reference values are established from lifetime exposure, C P Wild defines exposome as the collection and succession of individual and environmental exposures encountered by an individual during his/her lifetime [3]. This author proposes reconstructing an individual's given network of exposures in order to better understand the role of each exposure and thus generate research hypotheses as for the aetiology of diseases. In a similar way, Barabasi has constructed a "diseasome" in order to illustrate under the form of a network the environmental and social factors which would have a potential role in the origins of obesity [4]; and Goh et al. have explored a network of human diseases that implicate similar genetic mutations [5]. What all these analyses have in common is the research of similar characteristics for the understanding of mechanisms and the identification of key factors in the development and manifestation of diseases.

The exposome approach that we are in the process of developing has also an immediate application in health surveillance and the detection of emerging events for the RNV3P. This approach could be applied to a global analysis of health problems that would include in the RNV3P database other types of findings, variables and descriptors such as those used in the diseasome and exposome of the Wild type. A step or enterprise such as this greatly exceeds the simple framework, already complex in itself, of the RNV3P and necessitates going into and/or combining other databases.

Figure 1: Structure of the disease exposome.
Each tie represents a pathology for which the size is proportional to the number of related OHPs. The inter-connected links have at least one occupational exposure in common, such as the noxious substance. A) Exposome of disease categories of tumours, reported in the RNV3P database. B) Exposome of tumours. This category of diseases represents 4 sub-groups: malignant tumours, tumours of an unpredictable evolution, in-situ tumours and benign tumours. C) 38 diseases are distinguished amongst the malignant tumours, of which 22 isolated links do not have any noxious agents in common and the remaining 16 inter-connected links form a network as shown in fig. 1D. D) The tripartite network: disease–noxious agent–occupation is the origin of the exposome formed by 16 malignant tumours. The circles, squares and triangles represent malignant tumours, noxious agents and occupation, respectively, relative to each observation. The links between squares and triangles indicate noxious agents associated with the occupation for the OHP considered, and the links between squares and circles indicate noxious agents associated with the pathology.

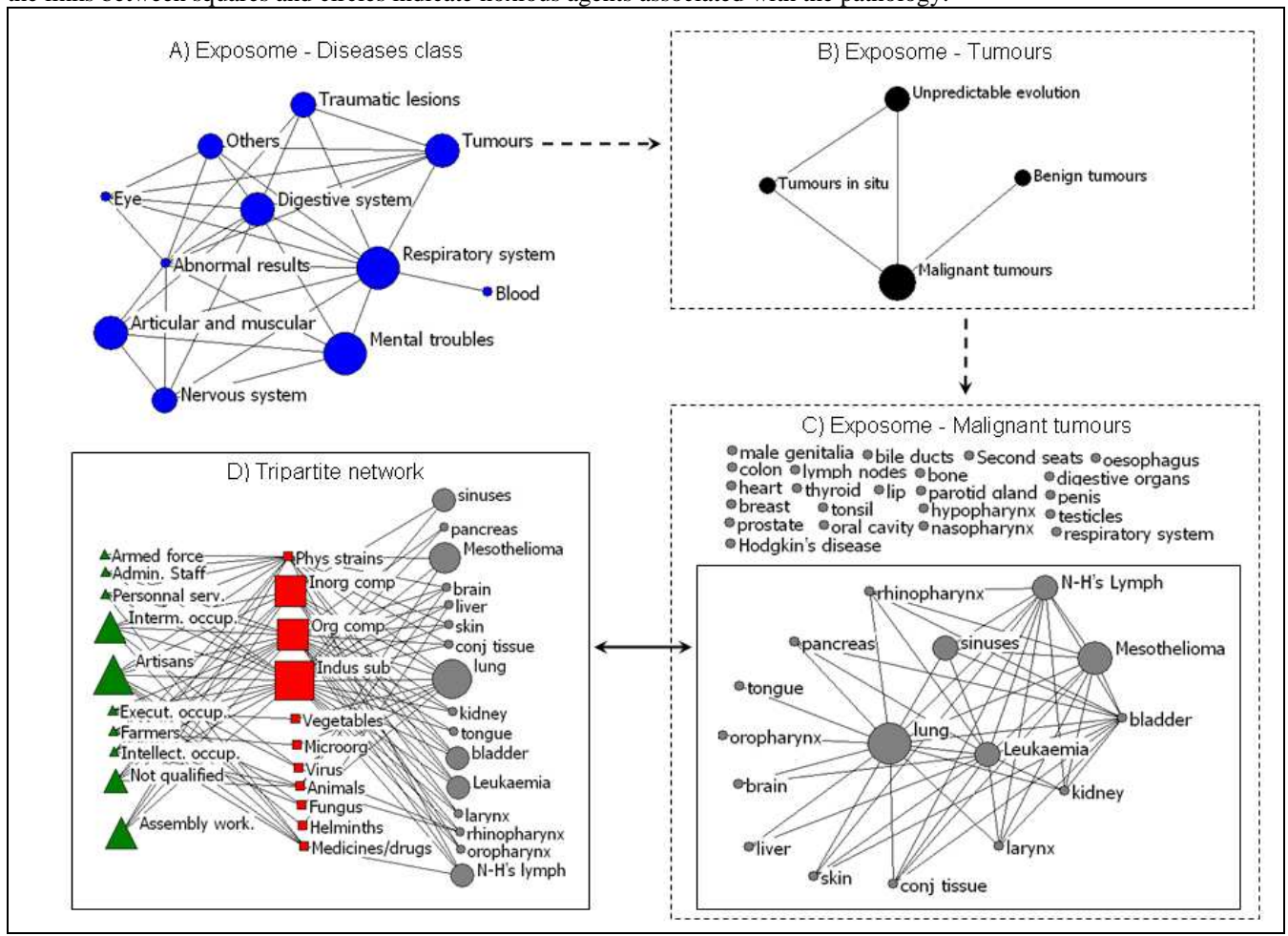